\documentstyle[12pt]{article}
\begin{document}
\title{Contribution of Inelastic Rescatterring to
 $B \rightarrow \pi \pi , K\bar{K}$ Decays}
\author{
{P. \.Zenczykowski}$^*$\\
\\
{\em Dept. of Theoretical Physics},
{\em Institute of Nuclear Physics}\\
{\em Radzikowskiego 152,
31-342 Krak\'ow, Poland}\\
}
\maketitle
\begin{abstract}
We discuss multichannel inelastic rescatterring effects in
$B$ decays into a pair $PP$ of pseudoscalar mesons 
($PP = \pi \pi$ or $K\bar{K}$). 
In agreement with short-distance models 
it is assumed that initially $B$ meson decays dominantly into jet-like states 
composed of two flying-apart low-mass resonances $M_1 M_2$
which rescatter into $PP$.
Since from all S-matrix elements $\langle i|S|PP\rangle$ involving $PP$
only some 
($i = M_1 M_2$) contribute to the final state rescatterring, the latter
is treated as a correction only. 
The rescatterring of resonance pair $M_1 M_2$  into the final $PP$ state
is assumed to proceed through Regge exchange.
Although effects due to a single intermediate state $M_1 M_2$
are small, it is shown that the combined effect of all such states  
should be large. 
In particular, amplitudes of $B$ decays into $K\bar{K}$ become significantly
larger than those 
estimated through short-distance penguin diagrams, to the point of
being comparable to the $B\rightarrow \pi \pi$ amplitudes.

\end{abstract}
\noindent PACS numbers: 13.25.Hw,11.80.Gw,12.40.Nn,12.15.Hh\\
$^*$ E-mail:
zenczyko@solaris.ifj.edu.pl
\begin{center}
REPORT \# 1849/PH, INP-Krak\'ow
\end{center}
\newpage

\section{Introduction}

Studies of CP-violation in $B$ decays must involve final state
interaction (FSI) effects. Unfortunately, a reliable estimate of such
effects is very hard to achieve. In the analyses of $B \rightarrow PP$ decays
($P$ - pseudoscalar meson) only some intermediate states, believed to provide 
nonnegligible contributions, are usually taken into account.
Many authors  restrict their studies to elastic rescatterring 
$P_1 P_2 \rightarrow P_1 P_2$ only. In Regge language 
this is described in terms of a Pomeron exchange.
Although in $PP \rightarrow PP$ quasi-elastic
rescatterring at $s=m^2_B$ contributions from 
other nonleading Regge exchanges  
are much smaller, they are not completely negligible and have been included
in various analyses.

The main problem, however, is posed by the sequence
$B \stackrel{weak}{\rightarrow} i \stackrel{FSI}{\rightarrow} PP $ 
involving {\em inelastic} rescatterring processes
$i \stackrel{FSI}{\rightarrow} PP$. 
Arguments have been given that it is these inelastic processes that actually 
constitute the main source of soft FSI phases 
\cite{Don96etal,SuzWolf99,Suzuki00,KamalLuo98}.
It has been also pointed out \cite{Kamal99} that
nonzero inelasticity strongly affects the extraction of FSI phases
in models based on quasi-elastic rescatterring. 
Thus, inelastic events affect model predictions even if rescatterring
is of quasi-elastic type only.

On the other hand, FSI phases are often attributed directly to short-distance
(SD) quark-line diagrams in the hope that this will take into
account all inelastic production phenomena.
This belief persists despite justified scepticism about the dominance of 
short-distance QCD in FSI of $B$ decays (see eg.\cite{Suzuki00}).
 In fact, it is known that the resulting prescription strongly violates 
 such tenets of
strong interactions as isospin symmetry \cite{GW97} (see also \cite{Wolf95}). 
The origin of the
problem pointed out in ref. \cite{GW97} 
is the lack of any correlation between the spectator quark and the
products of $b$ quark decay. 
By its very nature such correlation cannot be provided by SD
dynamics. What must be involved here is a long-distance (LD) mechanism
which ensures that quarks "know" about each other. 
Thus we are led to hadrons, hadron-level dynamics, and inelastic
rescatterring effects. 

In this paper we perform a simplified analysis of corrections 
which should be introduced by inelastic rescatterring 
into the SD-based description
of some nonleptonic decays of $B$ mesons.
Of course, any such analysis must be half-qualitative in nature, because - in
the presence of many decay channels - it is well 
beyond our ability to take them
accurately into account. Since it is believed that
$B$ decays into $\pi \pi$ and $K\bar{K}$ may 
provide some handle on the determination of angle
$\alpha $ of the unitarity triangle, we shall concentrate on these decays:
it is important to know the effects of FSI here.
As shown in ref. \cite{Zen99}, inclusion of coupled-channel quasi-elastic
effects ($\pi \pi \rightarrow K\bar{K}$) 
generates an effective long-distance penguin amplitude comparable
in size to the short-distance one.  
One may expect that inelastic channels will also contribute to this effect.

We start with an SD-based model of nonleptonic $B$ decays.  
On the basis of standard tree-dominated mechanism for 
these decays, enriched with the related and well-established models
of semileptonic $B$ decays, we qualitatively estimate the types and 
the number of states produced in the first stage of the nonleptonic decay.
As in other existing models, we take these states as composed of
two (flying apart) resonances $M_1 M_2$ (Section 2). These resonances are
assumed to rescatter into $PP$ through Regge exchange.

In order to provide the basis for an estimate of this rescatterring,
we recall how in a Regge picture the unitarity relation involving 
$M_1 M_2\rightarrow PP$ and other $i\rightarrow PP$ processes looks like.
This enables us to make a rough estimate as to what part of all inelastic
$i \rightarrow PP$ processes is due to
 the $M_1 M_2\rightarrow PP$ transitions and,
consequently, how much the situation deviates from the case of
(quasi-)elastic rescatterring.
Using a rough estimate for the contribution $|\langle M_1 M_2|S|PP\rangle|^2$ 
from an average single inelastic intermediate channel $M_1M_2$,
we estimate the number of inelastic channels involved (Section 3).

In Section 4
we analyse the behaviour of the $B \rightarrow \pi \pi$
and $B\rightarrow K\bar{K}$ amplitudes as a function of the number of 
intermediate states considered.
We show that
although effects due to each single intermediate state
are small, the combined effect of all intermediate states  
is large. 
In particular, amplitudes of decays into $K\bar{K}$ become significantly
larger than those estimated through short-distance penguin amplitudes, 
to the point of
being comparable to the $B\rightarrow \pi \pi$ amplitudes.

Our conclusions are given in Section 5.

\section{$B$ decays without FSI}
In short-distance approaches to nonleptonic weak $B$ decays, 
the relevant amplitudes are usually
expressed as sums of amplitudes corresponding to
different types of quark diagrams ($T$ - tree, $C$ - colour-suppressed, $E$ -
$W$-exchange, $P$ - penguin, $A$ - annihilation, $PA$ - penguin annihilation).
In this paper we concentrate on $\Delta S = \Delta C = 0$ decays of $B $ mesons 
which are initiated by a $b \rightarrow u\bar{u}d $ transition. In these decays 
one expects that the dominant contribution comes from the tree diagram $T$,
with the main corrections to it provided by the colour-suppressed and penguin
diagrams $C$ and $P$ \cite{GHLR95}.
Accordingly, 
neglecting contributions from other diagrams, the SD
amplitudes $\langle (P_1P_2)_I|w|B^0\rangle $ for
$B^0$ decays into a pair of octet pseudoscalar mesons 
$P_1P_2$ with total isospin $I$ are expressed 
in the SU(3) symmetry case as 
\begin{eqnarray}
\label{eq:qdBdecays}
\langle (\pi \pi )_2| w | B^0 \rangle & = & 
-\frac{1}{\sqrt{6}}(T+C) 
\nonumber \\
\langle (K \bar{K})_1| w | B^0 \rangle & = & 
-\frac{1}{2} P
\nonumber \\
\langle (\pi ^0 \eta _8)_1 | w | B^0 \rangle & = & 
-\frac{1}{\sqrt{6}} P
\nonumber \\
\langle (\pi \pi )_0 | w | B^0 \rangle & = & 
-\frac{1}{\sqrt{3}}(T-\frac{1}{2}C+\frac{3}{2}P) 
\nonumber \\
\langle (K \bar{K})_0| w | B^0 \rangle & = & 
\frac{1}{2}P 
\nonumber \\
\langle (\eta _8 \eta _8)_0 | w | B^0 \rangle & = &
\frac{1}{6} (C+P)
\end{eqnarray}
Inclusion of $\pi \eta_8$ and $\eta_8 \eta_8$ 
into our considerations is mandatory
if we want to maintain SU(3) symmetry \cite{Zen97}. 

From the above formulas one may find SD amplitudes ${\bf w}_{{\bf R},I}$ 
from state $|B^0 \rangle$ 
into states $\langle {\bf R}, I | $ of given
isospin $I$ belonging
to definite representations ${\bf R}$ of SU(3):
\begin{equation}
\label{eq:PPtoR}
{\bf w_R}_{,I} = {\bf O}_I {\bf w}_I
\end{equation}
with 
\begin{eqnarray}
{\bf w}_2^T&=& \langle (\pi \pi)_2 |w| B^0 \rangle \nonumber \\
{\bf w}_1^T&=&
[\langle (K \bar{K})_1| w | B^0 \rangle,
\langle (\pi ^0 \eta _8)_1 | w | B^0 \rangle]\nonumber \\
{\bf w}_0^T&=&[\langle (\pi \pi )_0 | w | B^0 \rangle
,\langle (K \bar{K})_0| w | B^0 \rangle,
\langle (\eta _8 \eta _8)_0 | w | B^0 \rangle]
\end{eqnarray}
and the matrices ${\bf O}_I$ given by:
\begin{equation}
{\bf O}_2=1,
\end{equation}
\begin{equation}
{\bf O}_1=\left[
\begin{array}{cc}
\sqrt{\frac{3}{5}} & \sqrt{\frac{2}{5}}\\
-\sqrt{\frac{2}{5}} &\sqrt{\frac{3}{5}}
\end{array}
\right]
\end{equation}
and 
\begin{equation}
{\bf O}_0 =\left[
\begin{array}{ccc}
-\frac{\sqrt{3}}{2\sqrt{2}} & \frac{1}{\sqrt{2}}& \frac{1}{2\sqrt{2}}\\
\frac{\sqrt{3}}{\sqrt{5}}&\frac{1}{\sqrt{5}}&\frac{1}{\sqrt{5}}\\
\frac{1}{2\sqrt{10}}&\frac{\sqrt{3}}{\sqrt{10}}&-\frac{3\sqrt{3}}{2\sqrt{10}}
\end{array}
\right]
\end{equation}
where rows correspond (from top to bottom) to $\bf 27$ for ${\bf O}_2$;
${\bf 8},{\bf 27}$ for ${\bf O}_1$; and
${\bf 1},{\bf 8},{\bf 27}$ for ${\bf O}_0$.

One expects that SD amplitudes $C$ and $P$ constitute a 10-20\%
correction \cite{GHLR95}. Indeed,
for the factor $r$ in 
the relation $C = T/(3r)$, the
short-distance QCD corrections give
the value of $ r = (c_1+c_2/3)/(3c_2+c_1) \approx -3$
($c_1 \approx 1.1 $, $c_2 \approx -0.25$ are Wilson coefficients), while
estimates in ref. \cite{G93} yield
$|P/T|$ in the range of $0.04-0.20$.

Dominance of $T$ amplitudes is expected to hold for other
decays initiated by $b \rightarrow u\bar{u}d$ as well.
For larger invariant masses of $d\bar{u}$ and $u\bar{d}$ systems
these quark-level states should hadronize mainly as 
$n \pi $ states \cite{Dobr89,Wolf91}.
It is the rescatterring from these states to the final PP state
that is of interest to us.
Since we want to estimate rescatterring at hadron level, we need to know
what hadronic states are produced in the first stage of the decay.
There are two groups of hadrons produced: one comes from the decay
$W^- \rightarrow d\bar{u}$, the other one originates from the
recombination of the $u$ quark with spectator $\bar{d}$ (Fig. 1).
The values of invariant mass squares $q^2 \equiv s_1$ and $s_2$ (see Fig. 1 for
definitions) are not large. The well-known probability distribution of 
$b$-quark decay
is 
\begin{equation}
\label{eq:bdecay}
v(q^2)=2 (1-q^2/m^2_b)^2 (1+2 q^2/m^2_b)
\end{equation}
 which falls for increasing
$q^2$, the average $s_1$ being $\overline{ s_1}  \approx 7~{\rm GeV^2}$.
The values of $s_2$ are smaller. Estimates obtained in various models
for semileptonic and 
nonleptonic decays 
yield  $\overline{ s_2}$  around 
 $1.5~{\rm GeV^2}$ or so, with the distribution
of $\sqrt{s_2}$ extending to around 2.0 or 2.5 GeV 
\cite{Dobr89,ACCMM,BKP,ISGW2,BDU,Battaglia}. 
When such a
low-mass quark-antiquark state hadronizes,
a resonance is produced.
In the ISGW2 model \cite{ISGW2} for semileptonic decays,
exclusive partial widths for the production of 
lowest-in-mass resonances have been predicted.
The resonances considered were the ground-state mesons  
(pseudoscalars $P$: $1 ^1S_0$ and vector mesons
$V$: $1 ^3S_1$), the P-wave mesons 
(tensor mesons $T$: $1 ^3P_2$, axial mesons $A$: $1 ^3P_1$ and 
$B$: $1 ^1P_1$, and scalar mesons 
$S$: $1 ^3P_0$), and the $2 ^1S_0$ and $2 ^3S_1$ states. The highest 
(nonstrange) resonance
mass explicitly considered in \cite{ISGW2} is below 1.5 GeV. 
The total partial width for the production of
all these
resonances is 5-6 times larger than the partial width for the production of a
pseudoscalar meson.  Thus, the average partial width into a resonance is
smaller by a factor of 0.7 or so 
than that for the production of a pseudoscalar meson.
The resonances explicitly considered in \cite{ISGW2}
do not saturate the inclusive decay rate which is still about two times larger
\cite{ISGW2}.
Thus, several other resonances of masses below 2.0 or 2.5 GeV 
should be added to the list given above. Assuming that the average
contribution
from each one is similar to that just estimated,
one expects that the number of types of "average" resonances produced
should be of the order of 15.

Similar or even larger number of resonance types is expected from the
hadronization of
the $d\bar{u}$ created from the $W$-boson.
Thus, in nonleptonic B decays, apart from the $PP$ state, many other resonance
pairs must be produced: $M_1 M_2$ = 
$VV$, ...,$PA$, $PB$, $VA$, $VB$, $VS$,...etc. 
Exact counting of the number of all these two-resonance
states is  not important for our purposes.  However, it is fairly easy to
give an estimate:  limiting oneself to
resonances of mass smaller than 2 GeV, this number
will definitely be greater than 10, probably of the order of a few tens.

Clearly, there is no hope that one can reliably calculate the contribution
from rescatterring into $PP$ from each of these intermediate states. However,
one may try to estimate their overall contribution in an average way.
For the first stage of the decay process, we will assume that the only
important amplitude is the tree amplitude $T$ and that this amplitude is
approximately the same for all intermediate states considered
(for low values of $s_1$ that we shall be concerned with later, this is
is indeed the case in Eq. (\ref{eq:bdecay})). 
Although
this may seem a very rough assumption for $s_2$,
we shall further see that our general results
should be fairly independent of it as long as there is a rather large
number of two-resonance states with production amplitudes scatterred around
the average.
The next question is how to describe transitions $M_1 M_2 \rightarrow PP$
(or vice versa) 
in an average way.  This is what we shall discuss in Section 3.

\section{Elastic scatterring 
and multiparticle production processes in PP interactions}

Elastic scatterring and multiparticle
production processes are related to each other
through unitarity of the S-matrix. Since $B$
has spin $J=0$, we shall work with ${\bf S}_{J=0}$ sector only. 
In the following we shall suppress the subscript $J$ and its value.
In refs. \cite{Zen99,Zen97} it was shown
that in the SU(3) symmetry case, with the effects of
coupled channels included,
one should work with states $(PP)_{\bf R}$ belonging 
to definite representations ${\bf R} ={\bf 1},{\bf 8},{\bf 27}$ of SU(3). 
The unitarity relation for $PP$ 
scattering in the $l=0$ partial wave and in SU(3)
representation ${\bf R}$ is:
\begin{equation}
\label{eq:unitarity}
|\langle (PP)_{\bf R}|S|(PP)_{\bf R}\rangle |^2+ \sum_{k_{\bf R}} 
|\langle (PP)_{\bf R}|S|k_{\bf R}\rangle |^2=1
\end{equation}
where, for given ${\bf R}$, $k_{\bf R}$ labels states different from
$(PP)_{\bf R}$. 
Matrix elements occuring in Eq. (\ref{eq:unitarity})
may be expressed in terms of Argand amplitudes $a$ as follows 
\begin{eqnarray}
\label{eq:elastic}
\langle (PP)_{\bf R}|S|(PP)_{\bf R}\rangle & =& 1 + 2 i a((PP)_{\bf R}) 
\nonumber \\
\langle (PP)_{\bf R}|S|k_{\bf R}\rangle &= & 2 i a(k_{\bf R})
\end{eqnarray}
Apart from the Pomeron, there are other Regge trajectories, whose exchange
in the $t$-channel contributes to quasi-elastic scatterring 
$(PP)_{\bf R} \rightarrow (PP)_{\bf R}$. 
When the leading non-Pomeron exchange-degenerate 
Regge trajectories  $\rho $, $f_2$, 
$\omega$, and $a_2$ and their SU(3)-symmetric partners are taken into account,
one obtains the $l=0$ partial wave amplitudes \cite{Zen99}:
\begin{eqnarray}
\label{eq:as}
a((PP)_{\bf 27})&=& \frac{1}{16 \pi} 
\left( i \tilde{P} +  \frac{2 \tilde{R} f(s)}{ s}
\right) \nonumber \\
a((PP)_{\bf 8})&=& \frac{1}{16 \pi}
\left( i \tilde{P} + \frac{\tilde{R} (-\frac{4}{3} f(s) + \frac{5}{3} g(s))}{ s}
\right) \nonumber \\
a((PP)_{\bf 1})&=& \frac{1}{16 \pi}
\left( i \tilde{P} + \frac{\tilde{R}
 (-\frac{2}{3} f(s) + \frac{16}{3} g(s))}{ s}
\right)
\end{eqnarray}
with
\begin{eqnarray}
f(s)&=& \frac{s^{\alpha (0)-1}}{ \ln (s)} \nonumber \\
g(s)&=& \frac{s^{\alpha (0)-1} \exp (-i \pi \alpha (0))}
{ \ln (s)}
\end{eqnarray}
where for the leading non-Pomeron Regge trajectory we use
\begin{equation}
\alpha (t) = \alpha (0) + \alpha ' t \approx 0.5 + t,
\end{equation}
ie. we have put $\alpha ' =1~{\rm GeV^{-2}}$ and, consequently, 
in Eqs (\ref{eq:as}) and further on both $s$ and $t$ are in ${\rm GeV^2}$.

The amplitudes $a((PP)_{\bf R})$ are
independent of isospin $I$ as shown in \cite{Zen99}, where
their sizes  at $s=m^2_B$ 
have also been estimated.
The SU(3)-symmetric Regge residue $\tilde{R}$  is fixed from experiment 
as \cite{Zen99}
\begin{equation}
\label{eq:Reggeon}
\tilde{R}/\alpha ' = -13.1{\rm mb~GeV^2} = -33.6
\end{equation}
while for the Pomeron one has \cite{Zen99}
\begin{equation}
\tilde{P}= 3.6~{\rm mb~Gev^2} = 9.25.
\end{equation}
Using the above values in Eqs (\ref{eq:as}) one finds
\begin{eqnarray}
\label{eq:asvalues}
a((PP)_{\bf 27})&=& -0.076 + 0.184 i \nonumber \\
a((PP)_{\bf 8})&=&  +0.019 + 0.217 i\nonumber \\
a((PP)_{\bf 1})&=& -0.076 + 0.291 i
\end{eqnarray}
while the leading non-Pomeron Regge contributions {\it alone} are:
\begin{eqnarray}
\label{eq:asReggevalues}
a((PP)_{\bf 27},Reg)&=& -0.076 \nonumber \\
a((PP)_{\bf 8},Reg)&=&  +0.019 + 0.033i\nonumber \\
a((PP)_{\bf 1},Reg)&=& -0.076 + 0.107 i
\end{eqnarray}
Contribution from elastic scatterring (the Pomeron in Regge language)
 is independent of ${\bf R}$:
\begin{equation}
\label{eq:Pomeron}
a((PP)_{\bf R},Pom)=\frac{i}{16 \pi}~ \tilde{P} = 0.184~ i
\end{equation}
(cf. $a = 0.17 ~i$ in ref. \cite{SuzWolf99}).
Omitting the Reggeon-Pomeron interference term,
the value of the contribution from the leading non-Pomeron Regge exchange
to the unitarity relation of 
Eq. (\ref{eq:unitarity}), after averaging over representations ${\bf R}$, is
equal to 
\begin{equation}
\label{eq:averagea}
|\langle PP|S_{Reg}|PP\rangle|^2 = 4 |a(PP;Reg)|^2 
\approx 4 |0.08|^2 =0.025
\end{equation}
Neglecting Reggeon contribution to 
$PP \rightarrow PP$ one obtains from Eq. (\ref{eq:unitarity}):
\begin{equation}
\label{eq:numericalunitarity}
\left( 1-\frac{\tilde{P}}{8\pi} \right)^2 + 
\sum_k |\langle PP | S | k \rangle |^2 =1
\end{equation}
One may conjecture that contributions to the above sum 
from Reggeon-exchange-induced processes
$PP \rightarrow M_1 M_2 $ (where $M_i$ denotes low-lying resonance) 
will be of size similar or smaller than 
$|\langle PP|S_{Reg}|PP\rangle|^2$. 
Thus, if all inelastic channels $k$ were two-resonance states $M_1 M_2$, 
one would obtain from Eq. (\ref{eq:numericalunitarity}) the number of
\begin{equation}
\label{eq:totalnumber}
n_{tot} =\sum_{M_1 M_2} |\langle PP | S | M_1 M_2 \rangle |^2/0.025 =
 \left( 1 - \left(1-\frac{\tilde{P}}{8\pi} \right)^2 \right)/0.025 \approx 25
\end{equation} 
as the number of states contributing to the unitarity relation.
This should be compared with the estimate of a few tens obtained in the
previous section for the number of two-resonance states produced in weak
decays of $B$ meson.
Of course, the estimate of Eq. (\ref{eq:totalnumber}) is probably too low:
 contributions from transitions $PP \rightarrow M_1 M_2$ 
for heavier resonances $M_i$ are 
likely to be smaller and the total number of states may be larger.
For example, with average value of $a$ going down by a factor of 0.6-0.7
from 0.08 to 0.05 
(Section 2), the above
estimate of the number
of average quasi-two-body states increases from 25 to over 60.
On the other hand, two-resonance states need not saturate
Eq.(\ref{eq:numericalunitarity}).
Thus, there are two important questions which should be answered: 
\begin{enumerate}
\item how many of states $k$ in the unitarity relation of
Eqs (\ref{eq:unitarity},\ref{eq:numericalunitarity})
are indeed of the form $M_1 M_2$, so that 
they can contribute to rescatterring in $B$ decays, and
\item can the effect of all of these states be described 
in some average way.
\end{enumerate}

At present, elastic and quasi-elastic 
contributions from long-distance FSI in $B$ decays are usually
evaluated using the old language of Regge theory. This language was 
used in the past also for the description of 
resonance and multiparticle
production processes that are of interest to us. 
In particular, the question of the 
build-up of elastic scatterring (Pomeron exchange) as a shadow of inelastic
multiparticle production processes (and thus the very content of the unitarity
relation of Eqs (\ref{eq:unitarity},\ref{eq:numericalunitarity}))
 was discussed extensively. 
Therefore,
we must recall the essential elements of an approach which dealt with that
problem. The approach, predominantly
occupied with the issue of the unitarization of dual models 
(for reviews see \cite{Chanrev}), 
was based on general properties of dual string models 
and on phenomenological analysis of resonance and multiparticle production data.

In this approach, multiparticle production processes occurring in
hadron-hadron collisions at high energies are pictured as
proceeding through the production of resonances or clusters
in a multiperipheral model (Fig. 2) with leading non-Pomeron
Reggeons exchanged in between the clusters. It is the {\it sum} 
over {\it all types} and
{\it numbers} of resonances produced in this way
that saturates the S-matrix unitarity relation
Eq. (\ref{eq:unitarity}).
For $s$ above the inelastic threshold but still small enough, 
one may limit oneself to the production
of just one pair of resonances $M_1 M_2$. With increasing energy $s$, the 
number
of resonances produced in a single collision increases on average. 
Although cross-section for a production of a particular number of resonances
goes down at sufficiently high energy, the sum of cross sections 
over all possible numbers of resonances remains
approximately constant. This constancy of the
total inelastic cross section is ensured by a sufficiently fast
increase in the number of all
possible quark-line diagrams, ie. in the number of all possible states $k$
(and ways in which they are produced) \cite{HuanLee73}.

In our case, at $s=m^2_B=28~{\rm ~GeV^2}$ the model \cite{CPTN75,CPT} 
predicts that 
states composed of just two resonances are produced
in the fraction of $f_{2M}\approx  50\% $ cases approximately. 
Further 35\% comes from production
of three resonances, etc. 
Although these numbers are obtained in \cite{CPTN75} for resonances of
any mass,
the main contribution comes
from production of resonances of  invariant mass squared smaller than $6 ~
{\rm GeV^2}$. (Contribution from the production of objects 
of mass $m_M$ is
suppressed as $(m_M^2)^{2(\alpha (0)-1)}$ for larger $m_M$ \cite{NSakai}).
The average mass of a resonance produced may be  estimated
in various ways to be around $\overline{ m_M} = 1.5$  ($1.7~{\rm
GeV}$ in ref. \cite{NSakai}), in good agreement 
with mass $\overline{m_M} \approx 1.3$ or so \cite{Battaglia},
expected for average $\sqrt{s_2}$  
in the SD-based models of weak decays. Contributions from
rescatterring of states with larger values of $s_1$ will be suppressed
because production of such states in $PP$ collisions is not likely.
Thus, in a rescatterring process $k \rightarrow PP $ 
one may expect that the dominant contribution
will indeed come from the rescatterring of states composed of two
low-mass resonances $M_1M_2$.
Translating the above expectation of a 50\% share of $M_1 M_2$ states in
the unitarity relation
of Eqs (\ref{eq:unitarity},\ref{eq:numericalunitarity})
into a number $n_{2M}$ of contributing channels 
which may connect to the state originally produced by the SD dynamics
(ie. $n_{2M}=f_{2M} n_{tot}$),
we conclude that this number should be around $n_{2M} =50\% \times 25 
\approx 12 $ for average 
$|a ( M_1 M_2)| \approx 0.08$ or $n_{2M} =50\% \times 60 
\approx 30 $
for average $|a ( M_1 M_2)| \approx 0.05$.
The latter estimate is probably more realistic since the average size
of a contribution from a single $M_1 M_2$ channel should diminish with
growing resonance masses.

\section{$B$ decays with FSI}
If one accepts that 
final state interactions cannot modify the probability of the original SD
weak decay, it follows that vector ${\bf W}$ representing the 
FSI-corrected amplitudes is related to vector ${\bf w}$ of the original 
SD amplitudes through \cite{SuzWolf99}
\begin{equation}
\label{eq:S1/2}
{\bf W} = {\bf S}^{1/2} {\bf w}
\end{equation}
Indeed, in the basis of ${\bf S}$-matrix eigenstates $|\lambda \rangle $ 
the above equation reduces to 
$W_{\lambda}=e^{i\delta_{\lambda}}w_{\lambda}$, ie. the
condition of unchanged probability 
($|W_{\lambda}|=|w_{\lambda}|$) admits Watson phases only.

The S-matrix may be written in terms of the matrix ${\bf A}$ of amplitudes $a$:
\begin{equation}
\label{eq:Amatrix}
{\bf S} = {\bf 1}+ 2 i {\bf A}
\end{equation}
We assume that we may
treat the FSI-induced corrections to the SD decay amplitudes in a perturbative
fashion. This is in agreement with the ideas of the dominance of SD dynamics.
If this assumption is incorrect, obtaining even half-quantitative predictions
will be almost impossible (cf. ref. \cite{SuzWolf99}).
Although this assumption may be questioned, it 
has an important advantage: one may study
what happens
when the number of contributing two-resonance intermediate states
 is increased to
its expected share ($f_{2M}=50\%$).
In agreement with the assumption of a perturbative treatment
of rescatterring (ie. small contribution from ${\bf A}$), we expand the square
root in ${\bf S}^{1/2}=(1 + 2 i {\bf A})^{1/2}$ and keep only the first term.
This leads to
\begin{equation}
\label{eq:Kmatrix}
{\bf W}\approx (1 + i {\bf A}){\bf w} = \frac{1+{\bf S}}{2} {\bf w}
\end{equation}
This is in fact the K-matrix prescription for the estimate of rescatterring
effects \cite{Kamal99}. In this prescription, final state interactions modify
probabilities of the SD amplitudes as can be seen explicitly
from Eq. (\ref{eq:Kmatrix})
written in the basis of S-matrix eigenstates.  
Our calculations will be based on Eq.(\ref{eq:Kmatrix}).

In previous sections we considered a simplified picture of contribution
from resonance pairs in which all amplitudes $a(M_1 M_2)$ were equal in absolute
magnitude while their phases were arbitrary. 
Indeed, when one row of 
the unitarity condition 
(ie. Eq. (\ref{eq:unitarity})) is discussed, no knowledge of phases is needed.
For the purpose of studies of CP-violation 
the question of phases is important, however. Therefore, we have to make
a very rough estimate of the FSI phases appearing 
in the strong rescatterring amplitudes 
$\langle (PP)_{\bf R} | A | (M_1 M_2)_{\bf R} \rangle$ in
\begin{equation}
\langle (PP)_{\bf R} | W | B \rangle = \langle (PP)_{\bf R} | w | B \rangle +
i \sum_{M_1 M_2} \langle (PP)_{\bf R} |  A | (M_1 M_2)_{\bf R} \rangle 
\langle (M_1 M_2)_{\bf R} | w | B \rangle .
\end{equation} 
Thanks to $CP$ invariance of strong interactions, 
the $\langle (PP)_{\bf R} | A | (M_1 M_2)_{\bf R} \rangle$ amplitudes are
symmetric (as is the ${S}$-matrix).
In order to estimate
them we have to recall what are the predictions of dual string models for
the production of two resonances in high energy $PP$ collisions.
In the Appendix of ref. \cite{CPT} it is shown that the dual string model
predicts that the amplitude for the $PP \rightarrow M_1 M_2$ production
through an uncrossed diagram of Fig. 3a
will pick up a rotating Regge phase resulting from the expression
\begin{equation}
(-s/(s_1 s_2))^{\alpha (t)}.
\end{equation} 
Similarly,
for the crossed diagrams of Fig. 3b one has to remove the
 $"-"$ sign in the above expression, ie. the amplitude is real.
Thus, the phase-generating factor differs from the familiar one in
$PP \rightarrow PP$ scatterring (ie. $(-s)^{\alpha (t)}$) 
only by a different scaling  factor  ($s_1 s_2$) 
in the denominator.
Such a dependence of Regge amplitudes on the masses of produced resonances 
has been confirmed in analyses of experimental data \cite{HRR73,IrvWor}.

At this point we have to take into account the fact that Regge amplitudes
describe scatterring of two colliding resonances $M_1$ and $ M_2$ 
in a state of definite momenta into a similar $PP$ state.
In particular, the produced $PP$ state is a 
superposition of partial waves, while we are interested in
the ${\bf S}_{J=0}$ sector of the $S$-matrix only.  
Restriction to the $J=0$ sector
is achieved by
integrating the rescatterring amplitudes $a(M_1 M_2)$ with 
$P_{l=0}(\cos \theta)$
(with $l$ being the angular momentum of the $PP$ pair)
over the allowed range of scatterring angle $\theta $,
or over the corresponding range of momentum transfer $t \in (t_{max},t_{min})$. 
Angular momentum conservation will then admit states $M_1M_2$ with
total angular momentum $J=0$ only. This will have no effect on the assumed
form of the SD decay amplitude since in the SD mechanism the 
decaying $b$ quark does not "know" about the
spectator quark, and, consequently, about the value of $J$ for the whole system.  

One calculates that for $s_1, s_2 \ll s $ the minimal value of $t$ is
$t_{min} \approx -s_1 s_2/s $, while for $t_{max}$ one may assume
$t_{max} \approx - \infty $.
Projecting the $M_1 M_2 \rightarrow PP$ Regge amplitude onto the $J=0$ sector,
ie. integrating Regge expressions over $t\in (t_{max},t_{min})$,
we obtain, up to a common overall normalization factor,
the following phase factors:
\begin{enumerate}
\item for the uncrossed diagram (Fig. 3a)
\begin{equation}
g_{inel}(z) = \frac{z^{\alpha (0)} 
\exp (-\ (\ln z - i \pi)/z - i \pi \alpha (0))}
{(\ln z - i \pi)}
\end{equation}
\item for the crossed diagram (Fig. 3b)
\begin{equation}
f_{inel}(z) = \frac{z^{\alpha (0)} \exp (- \ln (z)/z)}
{ \ln (z)}
\end{equation}
where $z\equiv s/(s_1 s_2)$.
\end{enumerate}
The overall normalization of contributions from amplitudes 
$M_1 M_2 \rightarrow PP$
was fixed in Section 3.

For the Regge description to be valid, the value of $s/(s_1 s_2)$ must be large.
With $s =m^2_B \approx 28 ~{\rm GeV^2}$, the product $s_1 s_2$ should 
not be greater
than $(s_1s_2)_{max} \approx 6~ {\rm GeV^4}$, possibly $9~{\rm GeV^4}$,
corresponding to the minimum value of $s/(s_1s_2)$ equal to $ 4 \pm 0.8 $. 
For $s_1$ equal to $s_2$,
this corresponds to the maximum value of
resonance mass being around 1.6-1.8 GeV. 
These are still reasonable numbers when compared with our
previous estimates of $\overline{ m_M} \approx 1.5$.
In Fig. 4 we show dependence of the phase of $g_{inel}(m^2_B/(s_1s_2)) $ 
on $s_1s_2$ for
$1 <  s_1 s_2 < 9 $. If one approximates $s_i$ by their average values
of around $(1.5 ~GeV)^2$, one finds that 
the average value of the product $s_1s_2$ 
is close to $5 ~GeV^4$. 
From Fig. 4 we see that for $s_1s_2 \approx 4.7 ~GeV^4$ the phase of 
$g_{inel}$ is zero.
Although the phase of $g_{inel}$ at smaller and larger values of $s_1s_2$
deviates from zero quite significantly, these deviations are on average of the
order of $20^o$ and, consequently, it is still meaningful to talk about
coherent superposition (ie. with approximately similar phases)
of Regge contributions from various intermediate states. 
  
Because the quark-line structure of $PP \rightarrow PP$
and $M_1 M_2 \rightarrow PP $ amplitudes is the same,
the eigenvalues with which the above phase factors enter into
expressions for $(M_1 M_2)_{\bf R} \rightarrow (PP)_{\bf R} $ in
definite SU(3) representations ${\bf R}$ are
the same as those in amplitudes $(PP)_{\bf R} \rightarrow (PP)_{\bf R}$.
The latter were calculated in ref. \cite{Zen99}.

On the basis of ref. \cite{Zen99} we have therefore
\begin{eqnarray}
\label{eq:deltas}
\delta ({\bf 27},M_1M_2) & = & \arg (+2 f_{inel}(s/(s_1 s_2))) \nonumber \\
\delta ({\bf 8},M_1M_2)  & = & \arg 
\left( -\frac{4}{3}f_{inel}(s/(s_1 s_2))+\frac{5}{3}g_{inel}(s/(s_1s_2))\right)
\nonumber \\
\delta ({\bf 1},M_1M_2)  & = & \arg 
\left( -\frac{2}{3}f_{inel}(s/(s_1 s_2))+\frac{16}{3}g_{inel}(s/(s_1s_2))\right)
\end{eqnarray}
The FSI-corrected amplitudes for $B$-meson decays into states in definite
representations ${\bf R}$ of SU(3) and with isospin $I$ are
\begin{equation}
{\bf W}_{{\bf R},I} = (1 + i a((PP)_{\bf R}) {\bf w}_{{\bf R},I}+ 
i \sum_{M_1M_2} e^{i \delta ({\bf R},M_1M_2)}~ 
{ |a((M_1M_2)_{\bf R})|} ~{\bf w}_{{\bf R},I}(M_1M_2)
\end{equation}
where ${\bf w}_{{\bf R},I}(M_1M_2)$ 
are SD weak decay amplitudes into $M_1M_2$, all assumed approximately
equal to ${\bf w}_{{\bf R},I}$ (Eq. (\ref{eq:PPtoR}), Section 2).
As in Section 3, we assume that {\it all} amplitudes $|a((M_1M_2)_{\bf R})|$ 
are equal to some
average value $\overline{a}$. 
 
Although the phase of $g_{inel}$ changes over the range of
corresponding $s_1s_2$, it is very instructive first
to approximate it everywhere
by a constant, namely its average evaluated at, 
say, $\overline{s_1s_2}\approx (\overline{m_M})^2
\approx ~ 4.7 GeV^4$, where $g_{inel}$ is real. The approximate reality of
average $g_{inel}$
is a consequence of the particular value of $s$ (being here equal to 
$m^2_B\approx 28~{\rm GeV^2}$) and not an $s$-independent feature
of the approach.  At $\overline{s_1s_2}=4.7~{\rm GeV^2}$ we get
\begin{eqnarray}
\overline{\delta ({\bf 1})}  &\approx & 0^o\nonumber \\
\overline{\delta ({\bf 8})}  &\approx & \pm 180^o\nonumber \\
\overline{\delta ({\bf 27})} &\approx & 0^o
\end{eqnarray}
The sum over all $n_{2M}$ two-resonance states will then yield
the contribution from inelastic rescatterring:
\begin{equation}
\label{eq:constphaseapprox}
{\bf W}_{{\bf R},I}(inel)= i~ n_{2M} e^{i \overline{\delta ({\bf R})}}~ 
\overline{ a} ~{\bf w}_{{\bf R},I}
\end{equation}
Note that for fixed $f_{2M}$ the average amplitude
\begin{equation}
\label{eq:constphaseapprox1}
\overline{a} =\sqrt{\frac{\left(1-\left(1-
\frac{\tilde{P}}{8\pi}\right)^2\right)f_{2M}}{4 n_{2M}}}
\end{equation}
is inversely proportional to $\sqrt{n_{2M}}$.
Thus, contribution from inelastic events in Eq. (\ref{eq:constphaseapprox}) 
is proportional to
$\sqrt{n_{2M}}$:
the smaller the value of $\overline{a}$,
the larger the summed contribution from all two-resonance states,
provided their contribution in the unitarity relation 
(Eq. (\ref{eq:unitarity})) 
is fixed by the same value of $f_{2M}$.
Obviously, 
this is a general feature of any perturbative treatment of rescatterring
contribution from several channels, the reason being the {\it linear} nature
of perturbatively treated FSI as compared to the {\it quadratic} nature of the
unitarity relation.
Thus, estimates of FSI effects given here are most likely estimates from
below: although 
amplitudes $a(M_1M_2)$ corresponding to rescatterring from states 
composed of resonances of larger masses 
are expected to be smaller,
their combined rescatterring effect should be relatively larger.

The amplitudes for 
$B$ decays into $(\pi \pi)_I$, $(K\bar{K})_I$
are calculated from the inverse of Eq. (\ref{eq:PPtoR}):
\begin{equation}
\label{eq:inverse}
{\bf W}_I = {\bf O}_I^T {\bf W}_{{\bf R},I}
\end{equation}
In Table 1 we give the values of $\langle (PP)_I|W| B^0 \rangle $
calculated from the above equation with the approximation of
average phase (Eq. (\ref{eq:constphaseapprox}))
for $PP = \pi \pi ,~K\bar{K}$ and $n_{2M}=12,~30$
($\overline{a} =0.08,0.05 $ respectively).
Predictions of the model without inelastic rescatterring, ie. with 
quasi-elastic FSI ($PP$ intermediate states) only, are also given
for comparison. Amplitudes are given in units of input tree amplitude $T$.

From Table 1 one can see that quasi-elastic FSI do not affect the decays
$B \rightarrow (\pi \pi)_I$ strongly. The amplitude moduli are somewhat smaller
than those without the FSI. 
This is due mainly to the $~1+i~[i\tilde{P}/(8\pi)]~$
factor originating from Pomeron exchange. Phase changes are small.
For the $B \rightarrow (K\bar{K})_0$ decays, quasi-elastic FSI affect
the amplitude significantly: the amplitude driven in SD-dynamics by the
penguin diagram (here vanishing)
receives contribution from the coupled-channel-generated long-distance
penguin (hereafter denoted $P_{LD}$) \cite{Zen99}.  
Using $|\langle (K\bar{K})_0|W|B^0 \rangle / \langle (\pi \pi)_0|W|B^0\rangle |
\approx \sqrt{3}P_{LD}/(2T)$ to estimate the effective LD penguin,
 the size of $P_{LD}$ 
is (as in ref. \cite {Zen99}) of the order of 5\% of the tree amplitude $T$, 
thus permitting significant
interference effects with the short-distance penguin amplitude
when the latter is taken from standard SD estimates.

The main results of this work are given in the last two columns of Table 1.
\begin{itemize}
\item One can see that the $B \rightarrow (\pi \pi)_I$ amplitudes, 
when compared with their estimates taking into account 
quasi-elastic FSI only,
increase
in absolute magnitude by a factor between 1 and 2.
This is due to the additional contribution coming
from rescatterring chain $B \rightarrow M_1M_2 \rightarrow \pi \pi$.
\item
An important change can be seen in phase sizes: they are now one 
order of magnitude larger than in
 the quasi-elastic case. The origin of this
effect is as follows. The $W$ amplitude is composed of two parts: one
(approximately real) is the SD amplitude $w$ weakly suppressed
by elastic (quasi-elastic) rescatterring, while the other contains
the contribution from the inelastic
$M_1M_2 \rightarrow \pi \pi$ rescatterring. The latter part, 
being proportional to
$i a(M_1M_2)w$, is mainly imaginary (for approximately real $a$).
With a large contribution from inelastic rescatterring, the resulting phase
must therefore be large.
\item
Finally, we see that the $B\rightarrow (K\bar{K})_0$ amplitude becomes
much larger than in the case of quasi-elastic rescatterring only, and 
is dominantly imaginary. 
Since the $B\rightarrow (K\bar{K})_0$  amplitude is fed
from no-hidden-strangeness states $(M_1 M_2)_{I=0}$, one might be tempted to
compare this with ref. \cite{Don96etal}. Namely, it was 
shown there for the case of
a simple two-channel S-matrix
that the phase of amplitude in channel 1 is large 
if the particle originally decays to channel 2. 
In our case, although a similar result holds, it is not general -
it depends on the value of $s$ at which amplitudes $a(M_1 M_2)$ 
are evaluated.
Note that inelastic rescatterring renders the $B \rightarrow (K\bar{K})_0$
amplitude larger than the SD penguin estimate: in fact, it is comparable 
to the $B \rightarrow (\pi \pi)_0$ amplitude.
\end{itemize}

In our approach
the full (FSI-corrected) 
amplitudes for $B \rightarrow (K\bar{K})_1$ decays vanish 
since: \\
1)we have assumed that only the tree amplitude $T$ is nonzero, and\\
2) the rescatterring from 
the isospin $|I=1,I_3=0\rangle $ state of $(M_1 M_2)_{I=1}$, 
which in principle might feed the final $(K\bar{K})_1$ channel, is zero
\footnote{Vanishing of the rescatterring contribution can be seen as follows.
The $I=1$ state of $M_1 M_2$ 
is antisymmetric (ie. of the form $(M^+M^--M^-M^+)/\sqrt{2}$), 
while the rescatterring contribution 
due to the uncrossed diagram of Fig. 3a is zero when evaluated
in between the antisymmetric state $|(M_1 M_2)_{I=1} \rangle $
and the symmetric state $|(K\bar{K})_1\rangle $
(for the definitions of states, see Eq. (1) of ref. \cite{Zen99}).
Rescatterring through diagram of Fig. 3b cannot change the type
of quarks and, consequently, cannot induce a transition from the
no-hidden-strangeness state  $|M_1 M_2 \rangle $ into $|K\bar{K}\rangle $.}.\\
The dependence of the LD rescatterring-induced effective penguin on
the isospin channel - with vanishing (large) effects in I=1 (0) channel -
should be compared with the SD mechanism which assigns the same size
and phase to penguin amplitude $P$ in both the $(K\bar{K})_0$ 
and $(K\bar{K})_1$ decay channels.
This is a general feature of long-distance dynamics:
the size and phase of a quark-level diagram depend on what isospin (SU(3))
amplitude it contributes to \cite{Zen97}.

Although the above expectation of FSI effect larger than that naively expected
is fairly general, we have to
analyse in some detail
our assumption of replacing the $s_1 s_2$-dependent phase of
$g_{inel}$ with an average (and vanishing) phase.
Such an assumption would be well justified if a large fraction of 
$M_1M_2$ states led to phases close to the average.
Since for a given value of the product 
$s_1s_2$ the phase is fixed, we need to know
the density of two-resonance states as 
a function of  $s_1  s_2$.
Regge models \cite{NSakai} predict that the dependence of $|a(M_1M_2)|^2 $
on $m^2_{M_i}$ is
proportional to $1/m^2_{M_i}$ for larger values of $m_{M_i}$. This fall of 
distribution for larger mass values should be clearly visible
already at the beginning of the
region where using Regge description becomes sensible, ie. at
$m^2_{M_i} \approx 4 ~GeV^2$ or so. The distribution
of $s_2$ in SD models vanishes even faster (it is fairly negligible above
$s_2 \approx (2.0-2.5 ~GeV)^2$). Since we want to estimate corrections
to SD models,
direct use of mass distributions generated in SD-models might seem to be
the simplest and most natural choice. Furthermore, as rescatterring of
states with larger values of $s_1$ is suppressed by the small size of the 
contribution from transitions $M_1M_2 \rightarrow PP$, one might use
the SD distribution of $s_2$ as that of $s_1$ as well.
The problem is, however, that
a significant part of the SD distribution of $s_2$ (or $s_1$) corresponds to 
$s_2 ~({\rm or}~ s_1)<1~{\rm GeV^2}$. On the other hand, Regge amplitudes
assume the form of 
$(s/(s_1 s_2))^{\alpha (t)}$ only for $s_1, s_2 >1~{\rm Gev^2}$.
For $s_{1(2)}<1~{\rm GeV^2}$, one should replace $s_{1(2)}$
with $(\alpha ')^{-1}=1~{\rm GeV^2}$. Thus, if
Regge phases are to be reasonably evaluated
we must use an appropriately modified distribution of $s_2$.
We model this situation in the simplest possible way: by 
assuming that the distribution of $m_{M_k}=\sqrt{s_k}$ vanishes below
$1~{\rm GeV}$ and above $2.25~{\rm GeV}$,
while in between these values it is given by
\begin{equation}
\label{eq:mMdistr}
\rho (m_M)= 2.88-1.28 m_M
\end{equation}
This yields the average value of "effective" $m_M$ equal to $1.42$, which
is also the value obtained from the ACCMM distribution of $\sqrt{s_2}$ (see
eg. ref. \cite{Battaglia}) if contributions from $m_M<1~{\rm GeV}$
are replaced with contributions at $m_M=1~{\rm GeV}$. Using $\rho (m_M)$
as in Eq. (\ref{eq:mMdistr}),
it is straightforward to evaluate the distribution $\rho _2$ of $s_1 s_2$.
This distribution (Fig. 5) is peaked at $s_1s_2 \approx 2.2 ~{\rm GeV^4}$, 
its median being at
$3.6~{\rm GeV^4}$, and the average $\overline{s_1 s_2}$ 
at $4.4~{\rm GeV^4}$. The
tail above $9~{\rm GeV^4}$ contributes a few percent only.
Thus, the assumption of $\overline{s_1s_2} \approx 4.7 ~{\rm GeV^4}$
used in our previous discussion appears quite reasonable.

With two-resonance states $k$  
spread in $s_1 s_2$ according to  distribution $\rho _2$
(larger values of $k=1,..,n_{2M}$ correspond to states with 
appropriately larger values of $s_1s_2$),
it is simple to analyse the predictions of the model
numerically.
We are interested in the question how the FSI effects change
when heavier and heavier intermediate states $k$ are included.
For decays into
states in definite representations of SU(3), the amplitudes of interest to us
 are therefore:
\begin{equation}
{\bf W}_{{\bf R},I}(n) = (1 + i a((PP)_{\bf R}) {\bf w}_{{\bf R},I}+ 
i \sum_{k=1}^n e^{i \delta ({\bf R},s_1s_2)}~ 
\overline{ a} ~{\bf w}_{{\bf R},I}
\end{equation}
where $n$ is the number of intermediate inelastic states considered
($0\le n \le n_{2M}$) and $\delta ({\bf R},s_1s_2)$ are given in 
Eq. (\ref{eq:deltas}).

In Fig. 6 we present predictions of the $n$-dependent version of
Eq. (\ref{eq:inverse}) 
 for the absolute values and
phases of the amplitudes of $B$ decays
into $(\pi \pi)_0$ and $(K\bar{K})_0$ states.
These predictions are given as a function
of the number $n$ of intermediate states considered.
We show the case with $n_{2M}=12$ (for $n_{2M}=30$ one obtains very similar 
plots with features discussed below being even more pronounced). 
The point most to the right in each plot
(ie. at $n=n_{2M}$) corresponds to a large value of $s_1 s_2$, where
Regge approximation breaks down. Consequently, this point should be
discarded.
One can see that for $B \rightarrow (\pi \pi)_0 $ 
the size of the amplitude does not depend very
strongly on $n$ and is close to the value of the input FSI-free amplitude 
$|\langle (\pi \pi)_0 |w|B^0\rangle|$, which is
$0.58$. 
Furthermore, the 
FSI-induced phase is still relatively small
(of the order of $-5^o$).
This cannot be said of the $B \rightarrow (K\bar{K})_0$ process.
Here, the absolute value of the amplitude grows fast with the increasing
number of intermediate states. In addition, already at $n=4$ or so,
 the phase becomes
close to $100^o$ , not far from the previous result of
$93^o$ (Table 1) obtained for constant phases. 
One can also see that when the number of
intermediate states approaches its maximum allowed value, the size of
the $B \rightarrow (K\bar{K})_0$ amplitude
becomes significant when compared to the
tree $B \rightarrow (\pi \pi)_0 $ amplitude.
As the number of intermediate states increases,
the long-distance-induced penguin amplitude
starts to dominate over the SD one (which
was estimated at 0.04 to 0.20 of the tree amplitude, \cite{G93}). 
Furthermore, assuming applicability of Eq. (\ref{eq:qdBdecays}), one estimates 
from Fig. 6 that at large $n$ 
the effective long-distance penguin amplitude $P_{LD}$ 
is around $0.6 ~{\rm to}~ 0.8$. This should be
compared with the input or effective $T$ amplitudes which are around $1.0$.
Thus, the long-distance effective penguin amplitudes become really 
large. 

Large rescatterring effects obtained above
stem from an {\it approximately coherent superposition of contributions
from several intermediate channels}.
One might argue that in the calculation of this paper 
sizes of individual  
rescatterring amplitudes $a$ (ie. contributions from 
individual intermediate channels) are overestimated. However, the trend of
the results expected for the case of smaller amplitudes $a$ may be seen 
from Table 1 and 
Eqs. (\ref{eq:constphaseapprox},\ref{eq:constphaseapprox1}).
The column of $n_{2M}=30$ corresponds to 
a smaller value of average amplitude $\bar{a}=0.05$. 
The connection existing between the values of $n_{2M}$ to $\bar{a}$ 
stems from the assumption that
at energy $s=m^2_B$  
a two-resonance state is produced
in approximately half of all inelastic $PP$ collisions.
As long as this fraction ($f_{2M}$) is kept constant and the perturbative
treatment of FSI is valid, smaller values of amplitudes $a$ 
- after summing over all
intermediate channels - result in
rescatterring corrections to weak decays larger than naively expected.
One may also try to estimate roughly 
the rescatterring contribution in a direct way
(ie. without using the value of $f_{2M}$) by
just adding rescatterring contributions
with amplitudes $a(M_1M_2)$ assumed to be of the order of $a(PP)$ or so
(in agreement with experiment). 
Then, for the number
of average intermediate states taken as equal to the number
 of final states in
standard models of SD decays,  ie. of the order of $10$ or more, one is
bound to obtain a large rescatterring effect.
Please note that 
our whole approach starts with the generally accepted features of the 
SD decay (and resonance production) mechanism and assumes that  
subsequent FSI may be treated perturbatively.
We have shown that even in this case the corrections tend to become large.
Of course, if they are too large, the whole perturbative scheme of their
estimation
(as well as the SD mechanism for the description of $B$ decays) 
ceases to be viable.

\section{Conclusions}
We have discussed multichannel inelastic rescatterring effects in
$B$ decays into $\pi \pi$ and $K \bar{K}$. Generally
accepted features of short-distance decay mechanism have been
assumed as part of our input. These assumptions
 included estimates of the types and
number of resonances produced in two-resonance states initiated by the
$b\rightarrow u\bar{u}d$ transition. Rescatterring of these two-resonance
states into the final state consisting of one pair $PP$ of pseudoscalar mesons
was evaluated under the assumption that such FSI may be treated perturbatively. 
The basis for this evaluation was provided by 
 existing knowledge about how the inelastic
multiparticle (resonance) production in $PP$ collisions is  
correlated with elastic $PP$ (Pomeron exchange) scatterring.
This knowledge permitted us to estimate that at $s=m^2_B$ 
$PP$ scatter into a two-resonance $M_1 M_2$ state
in approximately 50\% of cases.
Thus, the total size of rescatterring from $M_1 M_2$ to $PP$ was fixed.

Using the Regge model for the description of the $M_1 M_2 \rightarrow PP$ 
processes, we have shown
that the rescatterring contributions from the individual 
intermediate channels add approximately coherently.
As a result, the combined effect of rescatterring through
many two-resonance intermediate states was shown to be quite large.
This was demonstrated under the assumption that FSI may be
treated perturbatively. If that assumption is overoptimistic, 
reliable estimate of (presumably even larger) FSI effects will 
almost certainly be much more difficult. 

In our calculations the 
amplitude for the decays $B\rightarrow (K\bar{K})_0$ was
induced by LD rescatterring from no-hidden-strangeness isospin 0 states
produced via short-distance tree diagram. This FSI-induced amplitude
was shown to be larger than its
short-distance penguin counterpart. The phase of the LD amplitude
was estimated to be around $100^o$.

\section{Acknowledgements}
The author would like to thank Prof. P. Singer and 
the Department of Physics at the Israel Institute of Technology, Haifa, 
where some ideas of this paper have originated, 
for their hospitality.

\newpage
\noindent
FIGURE AND TABLE CAPTIONS

\noindent
Fig.1\\
Tree diagram ($T$) for $B$ decay

\noindent
Fig.2\\
Multiperipheral production of multi-resonance state $|M_1 M_2 ... M_n
\rangle $ through Regge exchanges $R_k$

\noindent
Fig.3\\
Quark-line diagrams for production of two-resonance 
state $|M_1M_2\rangle $:
(a) uncrossed Reggeon exchange, (b) crossed Reggeon exchange

\noindent
Fig.4\\
Dependence of $g_{inel}(m^2_B/(s_1s_2))$ phase on $s_1s_2$

\noindent
Fig.5\\
Distribution of $s_1s_2$ in model defined in text

\noindent
Fig.6\\
Dependence of FSI-induced effects on the number of intermediate
channels included.
\\
\\

\noindent
Table 1.\\
Effects of inelastic rescatterring on $\langle (PP)_I|W|B^0\rangle $ 
amplitudes in average phase approximation. Amplitudes are in units of 
input tree amplitude $T$.

\newpage
Table 1
\\
\\
\begin{tabular}{l c|c|c|c|c}
\hline
$(PP)_I$ & amplitude  & no FSI &quasi-elastic FSI & 
\multicolumn{2}{c}{inelastic FSI} \\
      & modulus/phase &        &                  & $n_{2M}=12$ & $n_{2M}=30$\\
\hline
$(\pi \pi)_2$ & $|W|$ & $0.41$ &   $0.33$ &  $0.49$ & $0.67$ \\
     & $\arg (W/w)$ & $0^o$  & $-5.3^o$ &  $47^o$ & $60^o$ \\
\hline
$(\pi \pi)_0$ & $|W|$ & $0.58$ &   $0.44$ &  $0.44$ & $0.61$ \\
     & $\arg (W/w)$ & $0^o$  & $-1.4^o$ & $-16^o$ & $-17^o$ \\
\hline
$(K\bar{K})_0$ & $|W|$& $0$    &  $0.028$ & $0.36$  &  $0.57$ \\
     & $\arg (W)$ &      &  $223^o$ &  $93^o$ &  $93^o$ \\
\hline
\end{tabular}


\newpage
\begin{thebibliography}{99}

\bibitem{Don96etal} J. F. Donoghue et al., Phys. Rev. Lett. 77, 2178 (1996).
\bibitem{SuzWolf99} M. Suzuki and L. Wolfenstein, Phys. Rev. D60:074019 (1999).
\bibitem{Suzuki00} M. Suzuki, {\em Final State Interactions in Heavy Hadron
Decay}, talk given at 3rd International Conference on B Physics and 
CP Violation, Taipei, Taiwan, 3-7 Dec 1999, hep-ph/0001170.
\bibitem{KamalLuo98} A. N. Kamal and C. W. Luo, Phys. Rev. D57, 4275 (1998).
\bibitem{Kamal99} A. N. Kamal, Phys. Rev. D60:094018 (1999).
\bibitem{GW97} J.-M. G\'{e}rard, J. Weyers, Eur. Phys. J. C7, 1 (1999).
\bibitem{Wolf95} L. Wolfenstein, Phys. Rev. D52, 537 (1995).
\bibitem{Zen99} P. \.Zenczykowski, Phys. Lett. B460, 390 (1999).
\bibitem{GHLR95} M. Gronau, O. F. Hernandez, D. London, J. L. Rosner,
Phys. Rev. D52, 6356 (1995).
\bibitem{Zen97} P. \.Zenczykowski, Acta Phys. Pol. B28, 1605 (1997).
\bibitem{G93} M. Gronau, Phys. Lett. B300, 163 (1993).
\bibitem{Dobr89} A. V. Dobrovolskaya, A. B. Kaidalov, K. A. Ter-Martirosyan,
and V. R. Zoller, Phys. Lett. B229, 293 (1989).
\bibitem{Wolf91} L. Wolfenstein, Phys. Rev. D43, 151 (1991).
\bibitem{ACCMM} G. Altarelli et al., Nucl. Phys. B208, 365 (1982).
\bibitem{BKP} V. Barger, C. S. Kim, and R. J. N. Phillips, Phys. Lett. B251, 629
(1990).
\bibitem{ISGW2} D. Scora and N. Isgur, Phys. Rev. D52, 2783 (1995).
\bibitem{BDU} I. Bigi, R. D. Dikeman, N. Uraltsev, Eur. Phys. J. C4, 453 (1998).
\bibitem{Battaglia} M. Battaglia, "Study of $b \rightarrow u l \bar{\nu}$
Decays with an Inclusive Generator" hep-ex/0002040.
\bibitem{Chanrev} Hong-Mo Chan, Sheung Tsun Tsou "Dual Unitarization: a New
Approach to Hadron Reactions", RL-76-080, Sep.1976, published in
Bielefeld Sum. Inst. 1976, 83; G. F. Chew, C. Rosenzweig, Phys. Rept. 41,
263 (1978).
\bibitem{HuanLee73} Huan Lee, Phys. Rev. Lett. 30, 719 (1973);
G. Veneziano, Phys. Lett. B43, 413 (1973).
\bibitem{CPTN75} Hong-Mo Chan, J. E. Paton, Sheung Tsun Tsou, Sing Wai Ng,
Nucl. Phys. B92, 13 (1975).
\bibitem{CPT} Hong-Mo Chan, J. E. Paton, Sheung Tsun Tsou, Nucl. Phys. B86, 479
(1975).
\bibitem{NSakai} N. Sakai, Nucl. Phys. B99, 167 (1975).
\bibitem{HRR73} P. Hoyer, R. G. Roberts, and D. P. Roy, Nucl. Phys. B56,
173 (1973)
\bibitem{IrvWor} A. C. Irving, R. P. Worden, Phys. Rept. 34, 117 (1977)

\end{thebibliography}
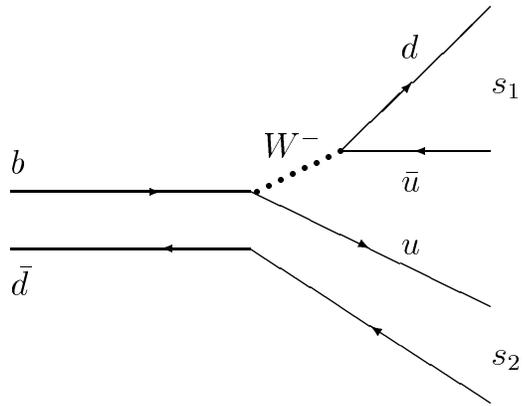
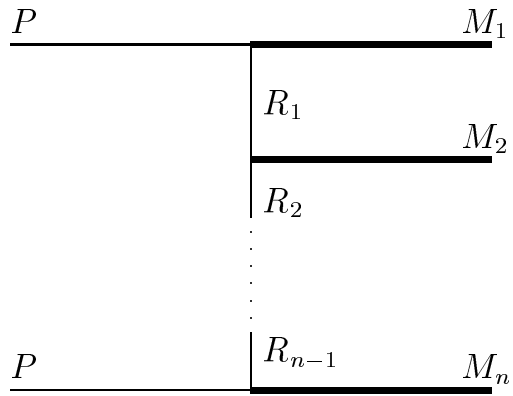
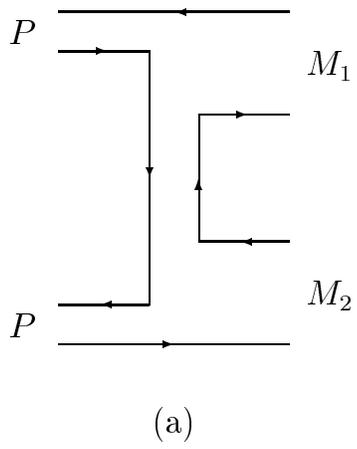
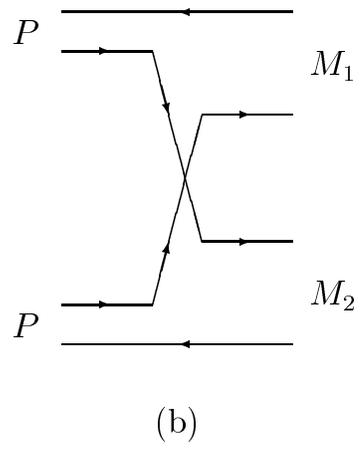
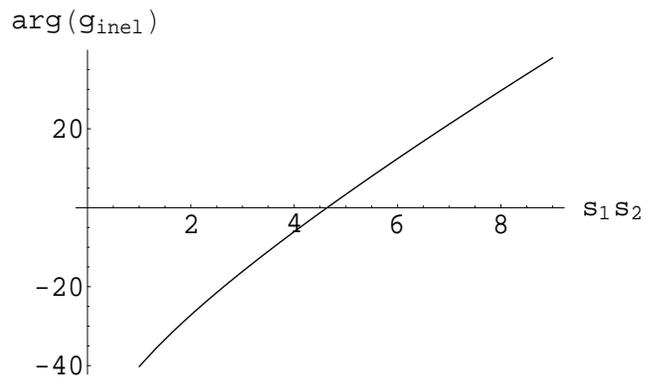
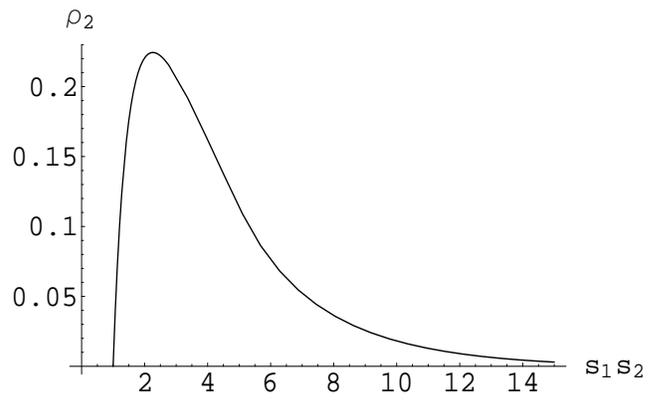
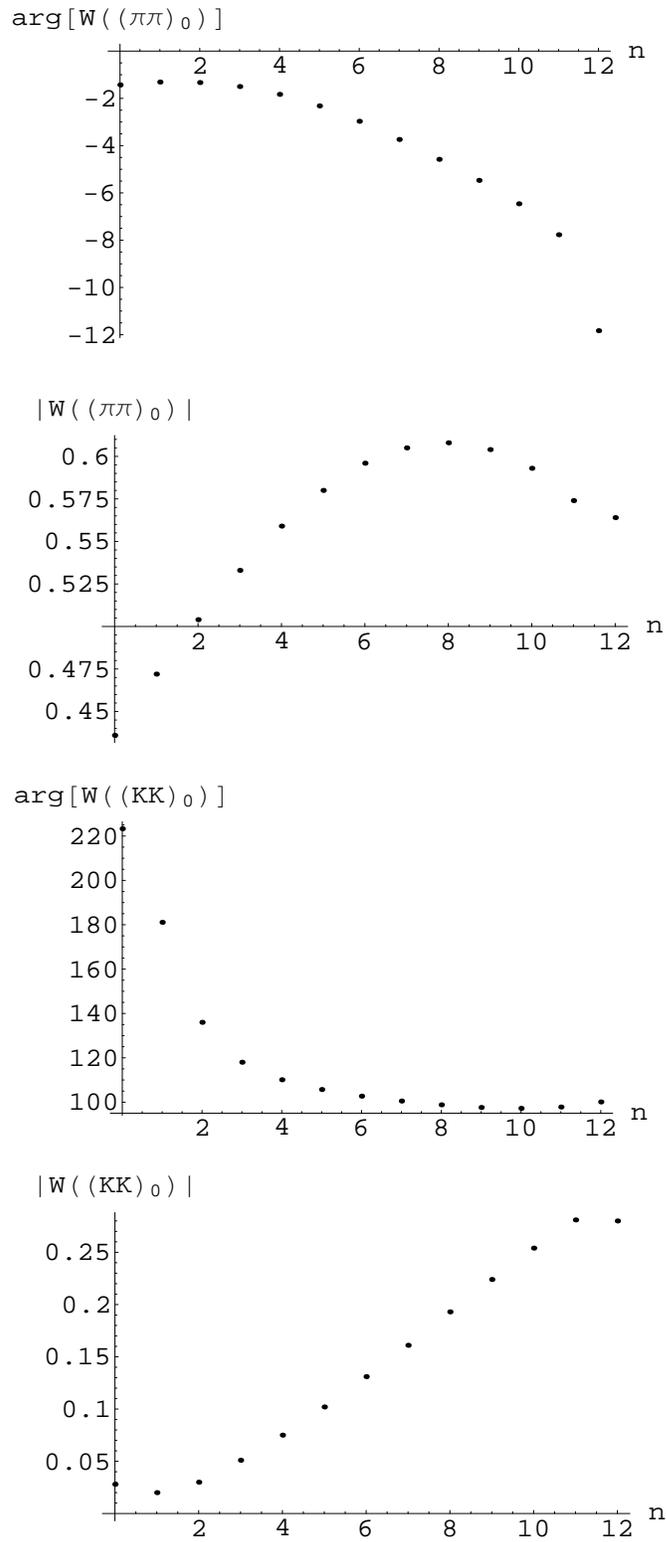
\end{document}